\documentclass[letterpaper]{article}

\usepackage{natbib,alifeconf}  
\usepackage{graphicx}

\usepackage{soul}
\usepackage{url}
\usepackage[hidelinks]{hyperref}
\usepackage[utf8]{inputenc}
\usepackage[small]{caption}
\usepackage{graphicx}
\usepackage{amsmath}
\usepackage{booktabs}
\usepackage{algorithm}
\usepackage{algorithmic}

\usepackage{color} 

\usepackage{hyperref}

\usepackage{amssymb}
\usepackage{graphicx}
\usepackage{enumerate}

\hypersetup{
    unicode=false,          
    pdftoolbar=true,        
    pdfmenubar=true,        
    pdffitwindow=false,     
    pdfstartview={FitH},    
    pdftitle={My title},    
    pdfauthor={Author},     
    pdfsubject={Subject},   
    pdfcreator={Creator},   
    pdfproducer={Producer}, 
    pdfkeywords={keywords}, 
    pdfnewwindow=true,      
    colorlinks=true,       
    linkcolor=blue,          
    citecolor=blue,        
    filecolor=magenta,      
    urlcolor=cyan           
}

\usepackage{todonotes}

\title{Exogenous Rewards for Promoting Cooperation in Scale-Free Networks}
\author{Theodor Cimpeanu$^1$,  The Anh Han$^1$,  Francisco C. Santos$^2$ \\
\mbox{}\\
$^1$School of Computing and Digital Technologies, Teesside  University\\
$^2$INESC-ID and Instituto Superior Tecnico, Universidade de Lisboa \\
Emails: \{T.Cimpeanu,T.Han\}@tees.ac.uk
} 

\begin{document}
\maketitle

\begin{abstract}

The design of mechanisms that encourage pro-social behaviours in populations of self-regarding agents is recognised as a major theoretical challenge within several areas of social, life and engineering sciences. When interference from external parties is considered, several heuristics have been identified as capable of engineering a desired collective behaviour at a minimal cost. However, these studies neglect the diverse nature of contexts and social structures that characterise real-world populations. Here we analyse the impact of diversity by means of scale-free interaction networks with high and low levels of clustering, and test various interference mechanisms using simulations of agents facing a cooperative dilemma. Our results show that interference on scale-free networks is not trivial and that distinct levels of clustering react differently to each interference mechanism. As such, we argue that no tailored response fits all scale-free networks and present which mechanisms are more efficient at fostering cooperation in both types of networks. Finally, we discuss the pitfalls of considering reckless interference mechanisms.

\end{abstract}


\section{Introduction}
\label{section:intro}
The problem of explaining collective behaviours among  self-interested individuals in evolving dynamical systems has fascinated researchers from many fields, and is  a well studied research topic in evolutionary game theory \citep{key:Hofbauer1998}. It can be found in a variety of real-world situations, ranging from ecosystems to human organisations and technological innovations and social networks~\citep{santos2006pnas,Sigmund2001PNAS,RaghunandanS12,hanAIES2019}. It has been also investigated in various Artificial Life systems such as  swarm-based systems and biologically inspired artificial social systems \citep{nitschke2005emergence,bonabeau}. 

 In this context, cooperation is typically assumed to emerge from the combined actions of individuals within the system. 
%
However, in many scenarios, such behaviours are advocated and promoted by an external party, which  is not part of the system, calling for a new set of heuristics capable of {\it engineering} a desired collective behaviour in a self-organised complex  system \cite{penn2010systems}. For instance, if one considers a near future, where hybrid societies comprising humans and machines shall prevail, it is important to identify the most effective incentives to be included to leveraging cooperation in such hybrid collectives \citep{paiva2018engineering}. In a different context, let us
%
 consider a wildlife management organisation (e.g., the WWF) that aims  to maintain a desired level of biodiversity in a particular region.
In order to do that, the organisation, not being  part of the region's eco-system, has to decide whether to modify the current population of some species, and if so, then when, and in what degree to \emph{interfere} in the eco-system (i.e., to modify the composition of the population)~\citep{levin2000multiple}.
Since a more impactful intervention typically implies larger costs in terms of human resources and equipment, the organisation has to achieve a balance between pregnant wildlife management and a low total investment cost. 
Moreover, due to the evolutionary dynamics of the eco-system (e.g., frequency and structure dependence)~\citep{santos2006pnas}, undesired behaviours can reoccur over time, for example when the interference was not sufficiently strong in the past. 
Given this, the decision-maker also has to take into account the fact that it will have to repeatedly interfere in the eco-system in order to sustain the level of biodiversity over time.
That is, it has to find an efficient  interference mechanism that leads to its  desired goals, while also minimising the its total cost.

This question has been studied previously in the context of populations distributed on regular graphs, namely the complete and the square lattice graphs \citep{han2018cost,hanIjcai2018}. In this type of network, every individual has the same  degree of connectivity (i.e. the number of neighbours). However, in social graphs and  real-world populations,  individuals typically have a diverse social connectivity \citep{albert2002statistical,santos2008social}.  
Hence, in this paper, we study  cost-effective interference in heterogeneous networks, namely different types of scale-free networks, which have been shown to well capture real-world networks (such as the  World Wide Web)~\citep{newman2018networks}.     
In particular, we consider populations of  individuals distributed in a scale-free network, who interact  with their neighbours via the one-shot Prisoner's Dilemma (PD),  where uncooperative behaviour is preferred over cooperation \citep{Sigmund2001PNAS,santos2006pnas}.
As an outsider decision-maker, we aim to promote cooperation by interfering in the system,  rewarding particular agents in the population at specific moments. 

The research question here is to identify when and how much to invest (on individuals distributed in a network) at each time step, in order to achieve our desired ratio of cooperation within the system such that the total cost of interference is minimised, taking into account the fact that individuals might have different levels of social  connectivity. For instance, we might wonder whether it is sufficient to focus the investment only on highly connected cooperators since they are more influential, thereby leading to cost-efficiency? Do we need to take into account  a neighbourhood's cooperativeness level which was shown to play an important role in square lattice networks \citep{hanIjcai2018}?  Also, when local information is not available and only global statistics can be used in the decision making process, how different are the results  in heterogeneous networks, in comparison to regular graphs?  

To answer these questions, this paper will systematically investigate different general classes or approaches of interference mechanisms, which are based {\it i)} on the global population statistics such as its current composition, {\it ii)} a node's social connectivity in the network and {\it iii)} the neighbourhood properties such as the local cooperativeness level.

Our results show that interference in a heterogeneous network exhibits a significantly more complex challenge (to be cost-effective while ensuring high levels of cooperation) and much richer nonlinear dynamic behaviours, compared to regular graphs. For instance, in both well-mixed and square lattice graphs, a greater per-individual investment  cost would ensure at least the same level of cooperation since it gives each cooperator a better fighting chance for survival against defectors. However, this is not the case in the context of heterogeneous networks as increasing the per-individual investment  cost could actually be detrimental for cooperation. 


The rest of the paper is structured as follows: the next section provides a brief overview of the related work,  which is followed by a detailed description of our model,  methods and its results. The paper ends with a final discussion.

\section{Related Work}
\label{section:relatedwork}

The problem of explaining the emergence and stability of cooperative behaviour  has been studied intensively in many fields,  from Social Sciences,  Economics, Physics to Multi-agent Systems and Artificial Life \citep{key:Hofbauer1998,nowak:2006bo,key:HanetalAlife,nitschke2005emergence}. Several mechanisms responsible for the evolution of cooperation have been identified, including  direct and indirect reciprocity~\citep{nowak:2005:nature}, kin and group selections~\citep{traulsen:2006:pnas}, network reciprocity \citep{santos:2005:prl,santos2006pnas}, punishment and rewarding~\citep{Sigmund2001PNAS}, and cognitive mechanisms \citep{key:hanetalAdaptiveBeh,key:HanetalAlife}. 
However, these mechanisms  do not consider how cooperation can be promoted by an external party. Instead, they are incorporated as part of individual  strategic behaviours, in order to study how they  evolve and whether their evolution promotes a better outcome for cooperative behaviour.
In contrast, our interference mechanisms are external, i.e. they are not incorporated into the individual strategy. 

In addition, the aim of our mechanisms is to minimise the cost of interference while guaranteeing  high levels of cooperation, contrary to past literature where the cost optimisation is often omitted. 
In this respect, our work is also different  from the modelling works of institutional incentives to  encourage  cooperation through costly  reward and punishment  \citep{sigmund2010social,vasconcelos2013bottom} as well as through enforcing agreements \citep{Han2017AAAI}.

Similarly, our work also differs from EGT  literature on optimal control in networked populations \citep{riehl2017towards,ramazi2015analysis}, where cost-efficiency is not considered. Instead, these works on controllability focus on identifying which individuals or nodes are the most important to control (i.e. where individuals can be assigned strategies as control inputs), for different population structures.


Closely related to the current work are the analyses on well-mixed populations (i.e. having a fully connected graph structure)~\citep{han2018cost} and on square-lattice structured populations \citep{hanIjcai2018}, which study cost-efficient interference on the aforementioned types of networks, respectively.
Moving to the more complex scenario of heterogeneous  networks where individuals might have different degrees of connectivity (i.e. the number of neighbours), an interference mechanism might  need to take this new dimension  into account to be cost-efficient. As shown below, cost-efficient interference mechanisms  that incorporate this information can outperform those who only consider global population statistics and neighbourhood cooperative properties as in previous works. 
		
Also related to current  work is the research of cooperation in social networks where  changes are initiated from inside the system~\citep{RaghunandanS12,FranksGJ13,FranksGA14}.
Among them, more relevant to our paper is the recent work by Franks \emph{et al.}~\citep{FranksGA14}, which has explored the use of influencers on complex networks. However, these influencers are also part of the system and thus, similar to the cases mentioned above, this work does not consider external interference mechanisms. Given this, it does not address similar decision-making problems that we examine here.

\section{Models and Methods}

\subsection{Prisoner's Dilemma on Scale Free Networks }\label{subsec:models}

We consider a population of agents on scale-free networks of contacts (SF NOCs)--- a  widely adopted heterogeneous population structure in population dynamics and evolutionary games (for a survey, see \citep{szabo2007evolutionary}). We focus our analysis on the efficiency of various interference mechanisms in spatial settings, adopting an agent-based model directly comparable with the setup of recent lab experiments on cooperation \citep{rand2014static}. 

Initially each agent is designated either as a cooperator (C) or defector (D) with equal probability. Agents' interaction is modelled  using the one-shot Prisoner's Dilemma game, where mutual cooperation (mutual defection) yields the reward $R$ (penalty $P$) and unilateral cooperation gives the cooperator  the sucker's payoff $S$ and the defector  the temptation $T$. As a popular interaction model of structured populations \citep{szabo2007evolutionary}, we adopt the following scaled payoff matrix of the PD: $T = b$, $R = 1$, $P = S = 0$. 
 (with $1 < b \leq 2$). We adopt a weak version of the Prisoner's Dilemma in spite of cooperator prevalence shown in previous works on scale-free networks of contacts \citep{santos2008social}, so as to have a direct link of comparison with studies on the effects of rewarding mechanisms in different types of networks \citep{hanIjcai2018}.

For SF networks with low clustering we adopt the famous Barab{\'a}si-Albert (BA) model \citep{albert2002statistical}. Starting from a complete graph of $m_0$ nodes, at every time-step one adds new node with $m \leq m_0$ edges linking to existing nodes, which are chosen with a probability that is proportional to the number of links that the existing nodes already have. The new node always connects to $m$ distinct nodes, and duplicate connections at each time step are not allowed. The average connectivity of the network is $z = 2m$.  

To obtain a SF network with high clustering, we resort to the Dorogovtsev-Mendes-Samukhin (DMS) model \citep{dorogovtsev2001size}. Similarly to the BA model, we also have growth, yet each new node attaches to both ends of a randomly chosen edge. As a result, we favor the creation of triangular relations between individuals, thereby greatly enhancing the clustering coefficient of the final network.  As in the BA model, the process of choosing the edge implicitly promotes the preferential choice of highly connected nodes, leading to the same degree distribution. The edges chosen at each time step are distinct and multiple connections between the same two nodes are not allowed. This network also has an average connectivity of $z = 2m$.  Both types of SF NOCs are pre-generated, before the strategies of players are designated and before the first generation commences playing. 

At each time step or generation, each agent plays the PD with its immediate  neighbours. The score for each agent is the sum of the payoffs in these encounters. Before the start of the next generation, the conditions of interference are checked for each agent and, if they qualify, the external decision maker increases their payoff. Multiple mechanisms (i.e. multiple conditions) can be active at once. At the start of the next generation, each agent's strategy is changed to that of its highest scored neighbour \citep{nowak1992evolutionary,szabo2007evolutionary}. Our analysis will be primarily based on this deterministic, standard evolutionary process  in order to focus on understanding the cost-efficiency of different interference mechanisms. 


We simulate this evolutionary process until a stationary state or a cyclic pattern is reached. The simulations  converge quickly, with the exception of some cyclic patterns which do eventually reach a stationary state. Because this work studies cost effective intervention, these rarely-occurring patterns which inherently invite very large total costs are escaped early by running simulations for only 75 generations, at which point the accumulated costs are excessive enough for this mechanism to not be of interest. Moreover, the results are averaged for the last 25 generations of the simulations for a clear and fair comparison (e.g. due to cyclic patterns). In order to improve accuracy related to the randomness of network topology in scale-free networks, each set of parameter values is ran on 10 different graphs for both types of SF NOCs. Furthermore, the results for each combination of network and parameter values are obtained from averaging 30 independent realisations. It is important to note that the distribution of cooperators and defectors on the network is different for every realisation. 

Note that we do not consider mutations or random explorations in this work. Thus, whenever the population reaches a homogeneous state (i.e. when the population consists of 100\% of agents adopting the same strategy), it will remain in that state regardless of interference. Hence, whenever detecting such a state, no further interference will be made. 

\subsection{Cost-Efficient Interference in Networks}\label{subsec:mechanisms}
As already stated, we aim to study how one can efficiently interfere in a structured population to achieve high levels of cooperation while minimising the cost of interference. An investment in a cooperator consists of a cost  $\theta > 0$ (to the external decision-maker/investor). This investment is added to the payoff of an agent if certain conditions are met. Each mechanism has different conditions for investment.  
In particular, we investigate whether global interference mechanisms (where investments are triggered based on network level information) or their local counterparts (where investments are based on local neighbourhood information) lead to successful behaviour with better cost efficiency.
To do so, we consider three main classes of  interference mechanisms based {\it i)} on the global composition of the population, {\it ii)} the node's connectivity in the network and {\it iii)} the neighbourhood cooperation level. 

\textbf{1. Population composition based (POP): } In this class of mechanisms the decision to interfere (i.e. to invest on all cooperators in the population) is based on the current composition of the population (we denote $x_C$ the  number of cooperators currently  in the population). Namely, they invest  when the number of cooperators in the population is below  a certain threshold, $p_C$ (i.e. $x_C \leq p_C$), for $1 \leq p_C \leq z$. They do not invest otherwise ($x_C > p_C$).  The value $p_C$ describes how widespread the defection  strategy should be to trigger the support of cooperators' survival against defectors.  

\noindent \textbf{2. Node Influence (NI): } For this mechanism, the decision to invest in a given cooperator is dependent on how influential its node is (i.e. how many connections end in that node). Whereas POP considered the composition of the population, NI looks at how connected a node is in the network. That is to say, the decision-maker invests in a cooperator node $C$ when the number of its immediate neighbours ($|k_C|$) divided by the maximum connectivity ($max |k_I|$) is above a threshold of influence $c_I$, for $0 \leq c_I \leq 1$. Otherwise, i.e. when $0 \leq c_I \leq \dfrac{|k_C|}{max |k_i|}$, no investment is made. The value $c_I$ describes how influential a cooperator node should be to trigger an investment into its survival.

\noindent \textbf{3. Local cooperation  based (LC): }  In this class of mechanisms, the decision to invest in a given cooperator is based on the cooperativeness level in that cooperator's neighbourhood.  Namely, the decision-maker invests in a cooperator when the number of its cooperative neighbours is below a certain threshold, $n_C$, for $0 \leq n_C \leq |k_C|$; otherwise, no investment is made.  By varying the local cooperation threshold $n_C$, we aim to provide an answer to the important question of how much cooperation is required in a neighbourhood before the investor can choose to withhold the intervention and save the interference cost and under which conditions this can happen. For instance, one can ask whether it is safe to withdraw action in a neighbourhood without affecting the outcome, therefore eliminating unnecessary interference.


Interestingly, these mechanisms require different levels of information which may or may not be readily available in the given network. In some cases, such as social networks, the connectivity (i.e. the number of friends) of a node is virtually free information which requires no effort on the part of the external decision maker to discern. On the other hand, other mechanisms such as POP, inherently require more information about the population and the level of cooperativeness in different parts of the network. POP is a broad mechanism which only requires knowledge about overall cooperativeness, but LC invites even more detailed observations, in order to determine the cooperativeness in each neighbourhood. Combining NI with LC generally does not require any more observation than LC by itself. Our study of neighbourhood based interference does not take into account the cost of gathering information, it is a direct comparison between perceived gains in cooperation and the associated per-individual cost of interference set out in the interference mechanisms.

\section{Results}

In contrast to the study on square lattice networks \citep{hanIjcai2018}, as detailed below for each interference mechanism, we found that performing cost-effective interventions on SF NOCs presents multiple concerns. In a square lattice population, more detailed observations resulted in more effective intervention with a better outcome. On the other hand, more knowledge about the population in SF NOCs simply reduces the risk of interfering to the detriment of cooperators. In other words, interfering in SF NOCs without adequate knowledge should be approached cautiously or it could act to the benefit of defectors. This issue is prevalent in the BA model and is not representative of the DMS model.

Positive interference in BA models broadly requires very high $\theta$ values (often orders of magnitude higher than similar mechanisms performed on square lattice populations) or a blanketing mechanism that targets all or almost all cooperators, even those which are not necessarily in danger of converting to D. Converging to 100\% C is very difficult unless both of these conditions are met and this introduces multiple concerns in the role of an exogenous interfering party. We avoid focusing on solutions where the per-generation cost is excessive, as it is unlikely for any institution to be able to produce such exorbitant sums in one generation, as required by these heterogeneous networks, instead we focus on effective intervention with manageable amounts of per-generation cost. In the following subsections we detail the results obtained for each interference mechanism.

\subsection{Population Based} 

\begin{figure}
\centering
\includegraphics[width=\linewidth]{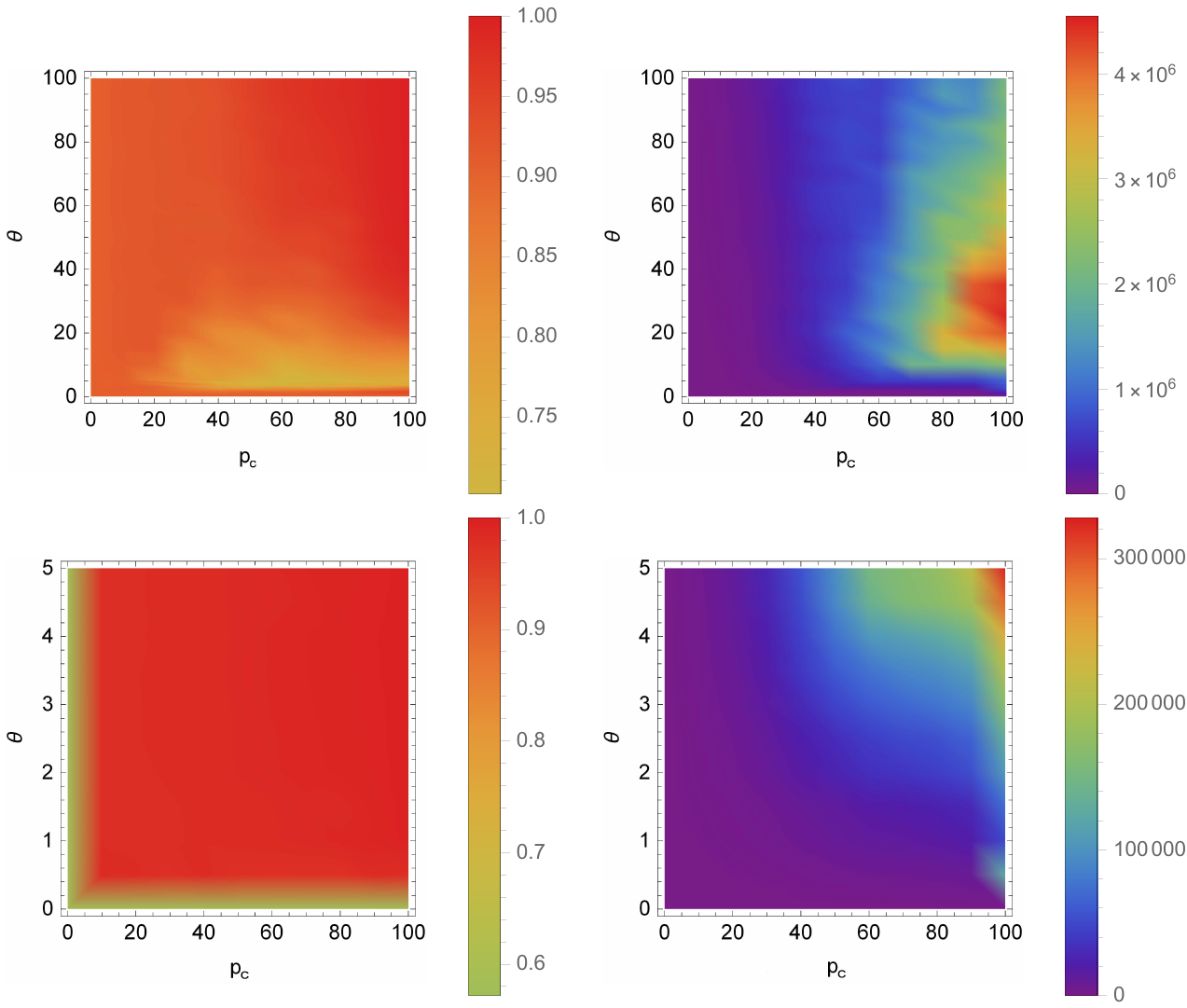}
\caption{\textbf{Population-based (POP) interference for BA model (top row) vs DMS model (bottom row),} for varying per-individual cost of investment $\theta$, as well as the threshold of population cooperation $p_C$. The left column reports the frequency of cooperation while the right one reports the total cost required. We note in particular, the significant difference in total cost between the two models. Parameters: $b = 1.8; \ n = 5000; \ z = 4$ (average node connectivity).}
\label{fig:pop}
\vspace{-0.6cm}
\end{figure}

We compare population-based interference mechanisms, i.e. POP, on the two different types of SF NOCs, the BA model and the DMS model, namely how efficient the mechanisms are at promoting cooperation with minimal total cost (See Figure ~\ref{fig:pop}).

For SF NOCs with a large clustering coefficient, we found that it is very easy to escape cyclic patterns and a minimal amount of interference, enabling the population to quickly converge to 100\% cooperation. Without any interference, the frequency of Cs is greatly dependent on the initial distribution of strategies in the network and there is a large probability that Ds will quickly overtake the Cs, if the oldest (i.e. the most connected) nodes are initially Ds. Conversely, applying even a minimal amount of interference to Cs, at any point in the rapid decline of C population, helps Cs in converging to 100\% of cooperation. Because of this, investing any more than minimal amounts of ($\approx \theta$), as well as interfering when the average cooperation is above 50\%, increases the total cost with little to no benefit to the frequency of Cs. Note that the results are consistent for a larger cost $\theta$. We plot up to $\theta \leq 5$ just for the sake of clear presentation.

\begin{figure}[t]
\centering
\includegraphics[width=\linewidth]{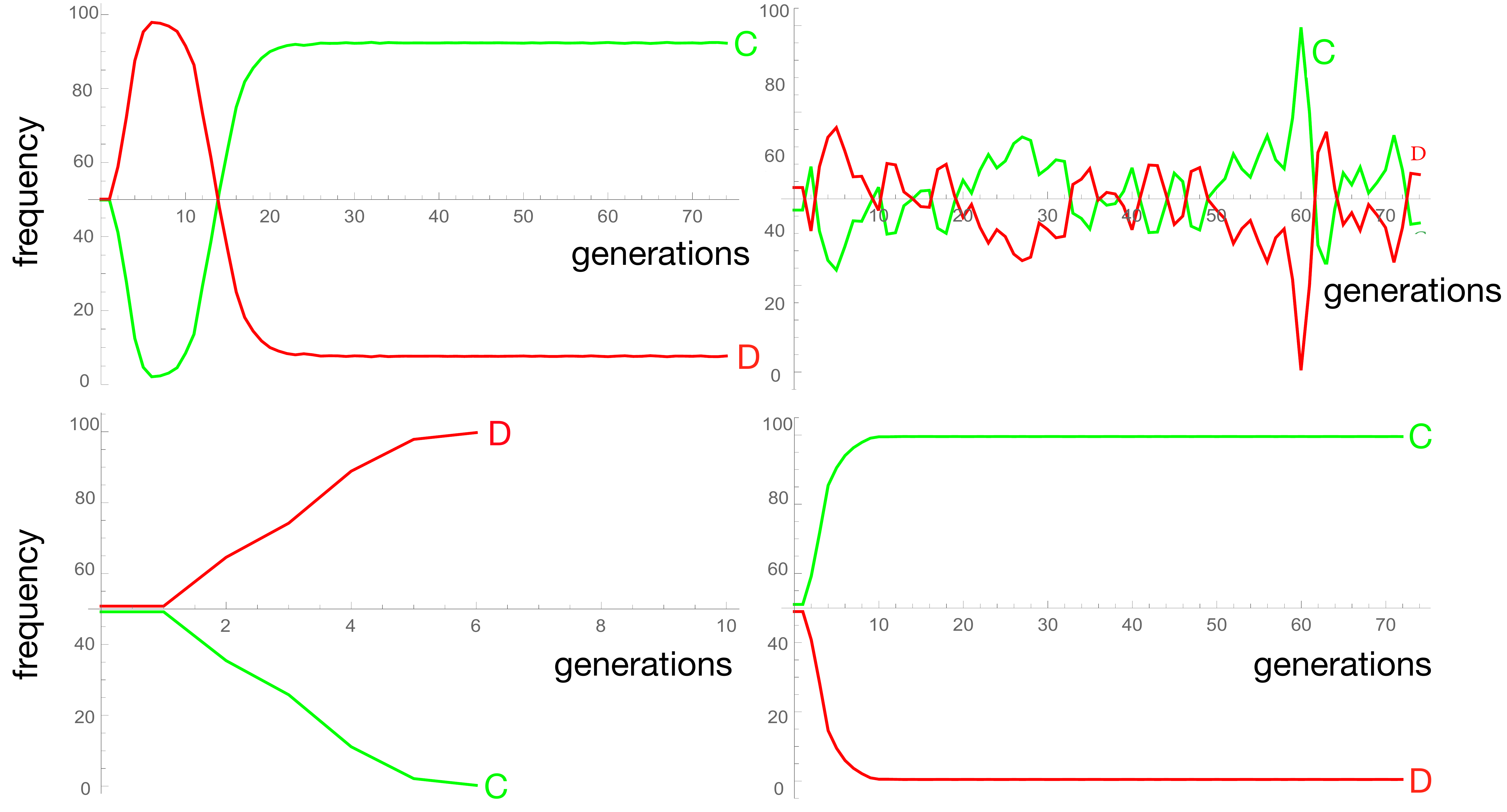}
\caption{\textbf{Evolution of cooperation: BA Model (top) vs DMS model (bottom),} for $\theta = 5, p_C = 80$. C and D frequencies are shown in green and red, respectively. The left column shows the network without interference, while the right one shows the same network after population-based (POP) interference. Other parameters: $b = 1.8; \ n = 5000; \ z = 4$ (average node connectivity).}
\label{fig:bad-interf}
\vspace{-0.3cm}
\end{figure}

In direct contrast with our findings for the DMS model, an external decision maker should only interfere in BA models with great care, as investing without discrimination could lead to a lower cooperation frequency when compared to no interference (See Figure ~\ref{fig:pop}). We observe that using certain values of $\theta$ negatively impacts average cooperation levels across a wide range of $p_{C}$ values. For these undesired values of $\theta$, cyclic patterns would form which ultimately help Ds by maintaining C players in clusters dominated by Ds (see Figure \ref{fig:bad-interf}). This type of negative impact occurs when the $\theta$ value is not high enough for Cs to be able to convert a cluster to cooperation, but not low enough as to let the Ds converge to 100\% D in that cluster. Many of these cyclic patterns eventually settle to 100\% C if the simulation is ran for a sufficient number of generations ($\approx 250$). We note that the accumulated cost of interference at the end of the long-lasting cyclic patterns is prohibitively large, which make such values of $\theta$ undesirable for an external decision maker with limited resources.

Positive interference in BA models can be achieved by selecting very low or extremely high values for $\theta$, with a high value for $p_C$. BA models converge to a high C frequency even without interference, so it is important to select a value for $p_C$ that will allow interference after the system has reached a stable state (typically $p_C > 90$). In terms of total cost, it is more efficient to select very low values of $\theta$, but the overall benefit to cooperation levels is much lower than with very high values of $\theta$. Therefore, it is up to the external decision maker to decide if the increase in cooperation is worth the higher cost in resources. 

\subsection{Node Influence}
\begin{figure}[t]
\centering
\includegraphics[width=\linewidth]{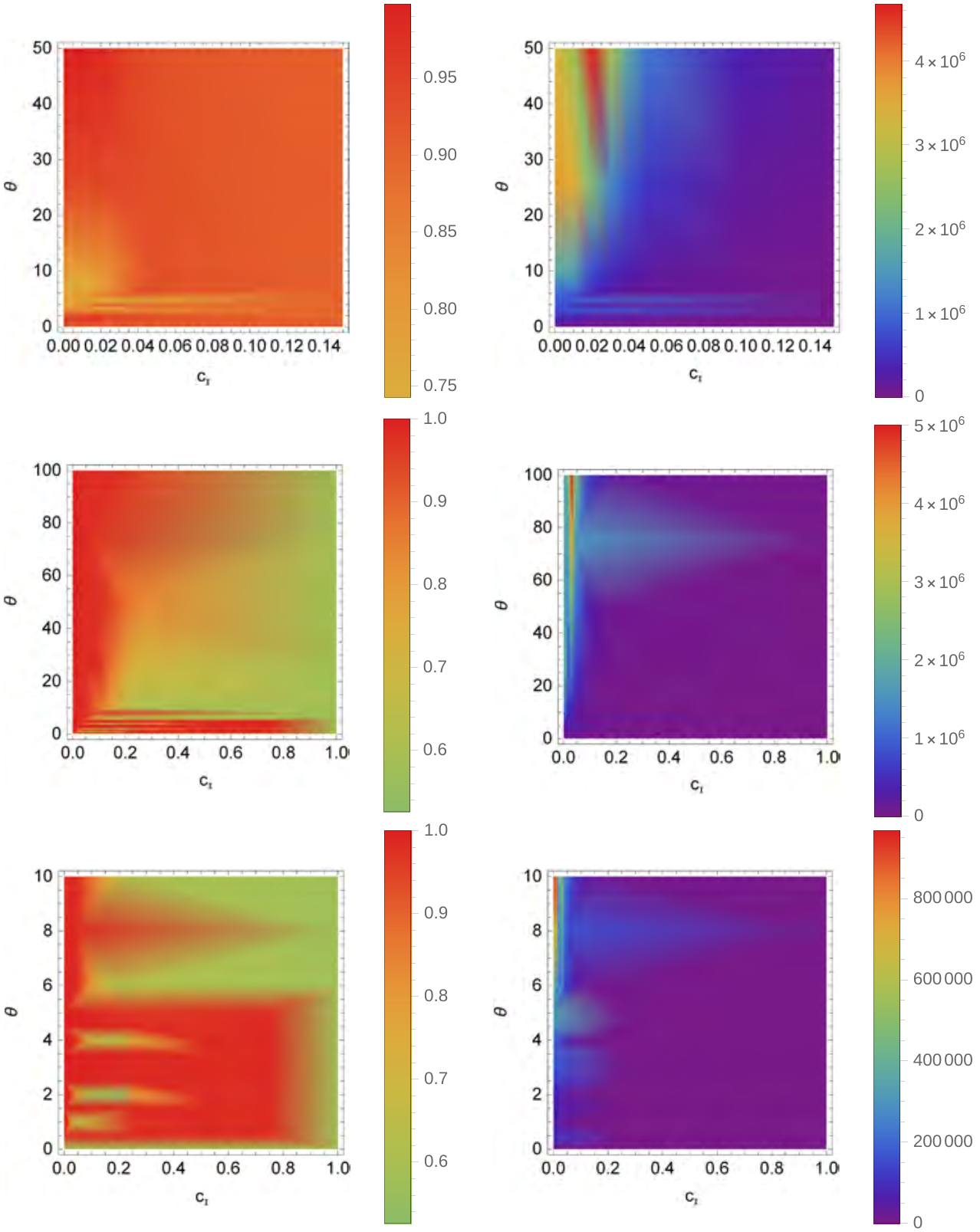}
\caption{\textbf{Node influence based (NI) interference for BA model (top row) vs DMS model (middle row) and detailed view of DMS model (bottom row),} for varying per-individual cost of investment $\theta$, as well as the threshold for cooperator influence $c_I$. The left column reports the frequency of cooperation while the right one reports the total cost required. The ranges for $\theta$ and $c_I$ are scaled for clear presentation. Parameters: $b = 1.8; \ n = 5000; \ z = 4$ (average node connectivity).}
\label{fig:neb-influence}
\vspace{-0.3cm}
\end{figure}

When an exogenous decision maker takes into account only how connected a node is, see Figure \ref{fig:neb-influence}, i.e. how influential it is in the network, it becomes very unlikely for interference to provide a meaningful improvement to levels of C in BA models. Very low values of $c_I$ ($\approx 0.1$), coupled with small to intermediate values of $\theta$ can cause a decrease in the average cooperation, forming previously discussed cyclic patterns (See Figure ~\ref{fig:bad-interf}). For all other values, cooperation seems to be very inert. This phenomenon can be explained by the fact that clusters have already been decided in favour of cooperators and all that remains of defectors survives in a stable state around non-influential nodes. By targeting only the most influential cooperators, the external decision maker can have no impact on the less connected nodes, which enable the survival of defectors. Therefore, only a blanketing mechanism at very low  $c_I$ can reach the lowly connected cooperators and produce an increase in the average cooperation. This type of blanketing mechanism with low $c_I$ quickly accumulates large amounts of total investment cost.

In SF networks with a high clustering coefficient, on the other hand, one can ensure convergence to 100\% C in a cost effective way by selecting intermediary values of $c_I$ (typically $\approx c_I = 0.6$) and low values for $\theta$. An interesting observation is that contrary to POP, interference does not mean that the system will converge to 100\% C. Anything more than minimal amounts of $c_I$ show no increase to average of cooperation except at extremely high $\theta$ values. In other words, it appears that D clusters are very difficult to shift after the initial distribution of strategies on the network and the amounts of fitness they acquire are almost impossible to match except for a very large individual investment into the oldest and, implicitly, most influential, nodes. Similarly to the effect of NI on the BA model, a blanketing mechanism encourages the formation of C dominated neighbourhoods, which in turn generates a greater fitness than anything but very large values of individual investment $\theta$. Interestingly, increasing individual investment at anything but low values of the influence threshold $c_I$ actually promotes defection by enabling the temporary survival of cooperators connected to defectors which are centers of hubs. This, in turn, allows defector hubs to convert any remaining cooperator hubs. It is important to note that the initial distribution of players in the hubs is ultimately what determines which way the network will converge, so this type of interference does not produce any decrease in cooperation, as is the case of the BA model.

\subsection{Local Cooperation}
\begin{figure}[t]
\centering
\includegraphics[width=\linewidth]{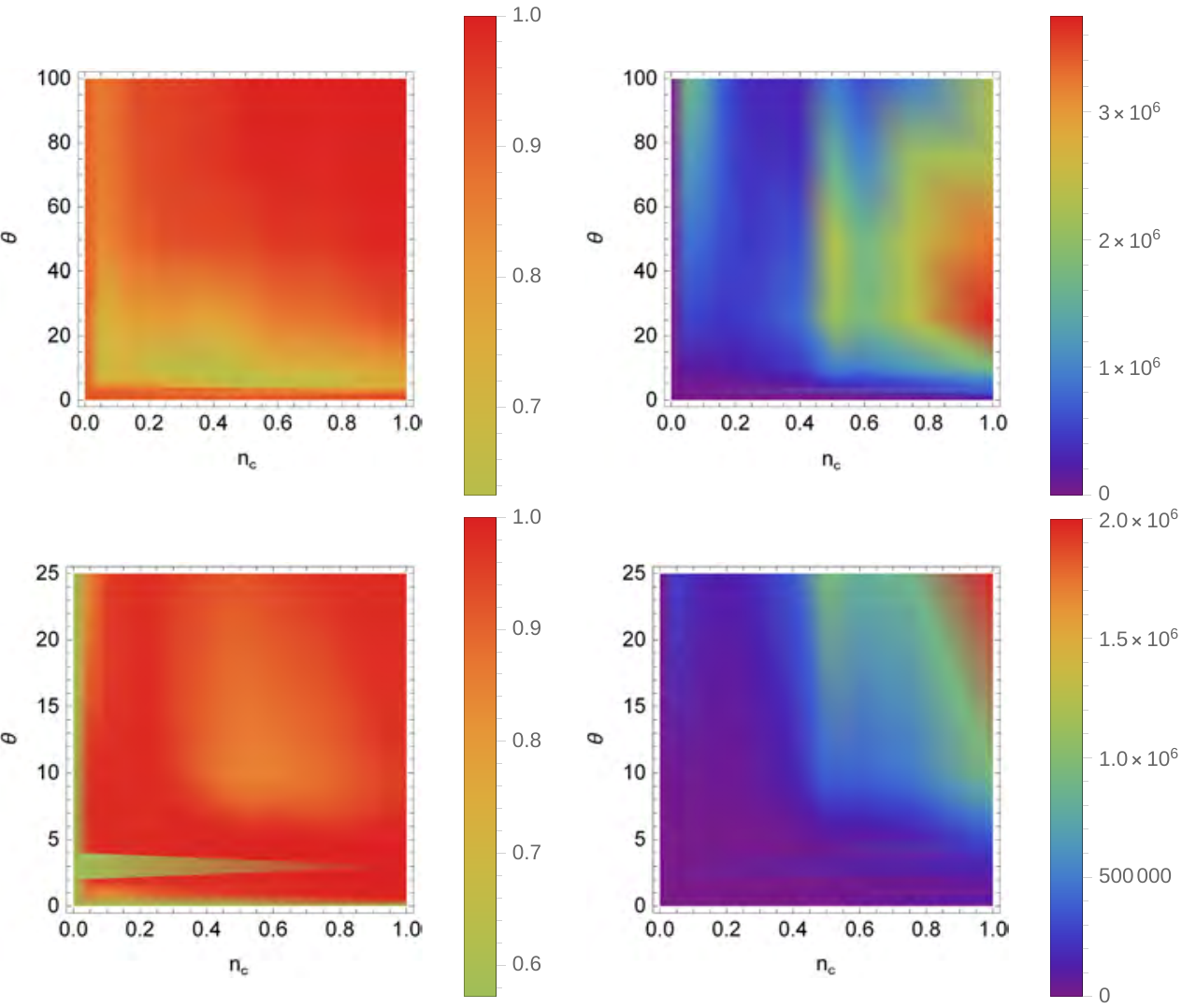}
\caption{\textbf{Local cooperation based interference for BA model (top row) vs DMS model (bottom row),} for varying per-individual cost of investment $\theta$, scaled for clear presentation, as well as the threshold of neighbourhood cooperation $n_C$. The left column reports the frequency of cooperation while the right one reports the total cost required. Parameters: $b = 1.8; \ n = 5000; \ z = 4$ (average node connectivity).}
\label{fig:lc-simple}
\vspace{-0.3cm}
\end{figure}

By the same token as earlier observations, interference on the BA model comes with the risk of reducing overall cooperation. What is more, LC based interference produces negative results for a wider range of parameters than any other mechanism (See Figure ~\ref{fig:lc-simple}). That notwithstanding, investing smartly using the LC mechanism can lead to 100\% cooperation, whereas the other mechanisms struggle. The key to such smart investments is in choosing a value for the threshold of local cooperation $n_C$ which approaches the upper limit ($n_C \to 1$), with individual investment $\theta$ values high enough to convert defectors situated in cooperator clusters. As the value of $n_C$ approaches $1$, redundant investment decreases. With higher values of $\theta$, the network converges more rapidly and therefore overall cost is reduced. Therefore, LC based interference can be regarded as the least risk averse, but potentially the most impactful given realistic values of per-individual investment $\theta$. 

With the exception of a small range of values for per-individual investment $\theta$, the LC interference mechanism achieves a very high average cooperation for the clustering network. This reinforces the assumption that interference at any point in the decline of cooperation is enough to shift the scales and enables cooperators to overtake the defectors. On the basis thereof, an external decision maker can reduce the costs of interference by selecting very low values for $n_C$ in combination with a high enough individual investment $\theta$. For intermediate values of the local cooperation threshold $n_C$ in combination with not high enough $\theta$ values, an interesting phenomenon is observed: the promotion of the survival of defectors by enabling the survival of cooperator nodes without actually giving them a chance of converting defectors in their neighbourhood, thereby allowing the defectors to exploit those cooperators. 

\subsection{Combining Node Influence and Local Cooperation}

\begin{figure}
\centering
\includegraphics[width=\linewidth]{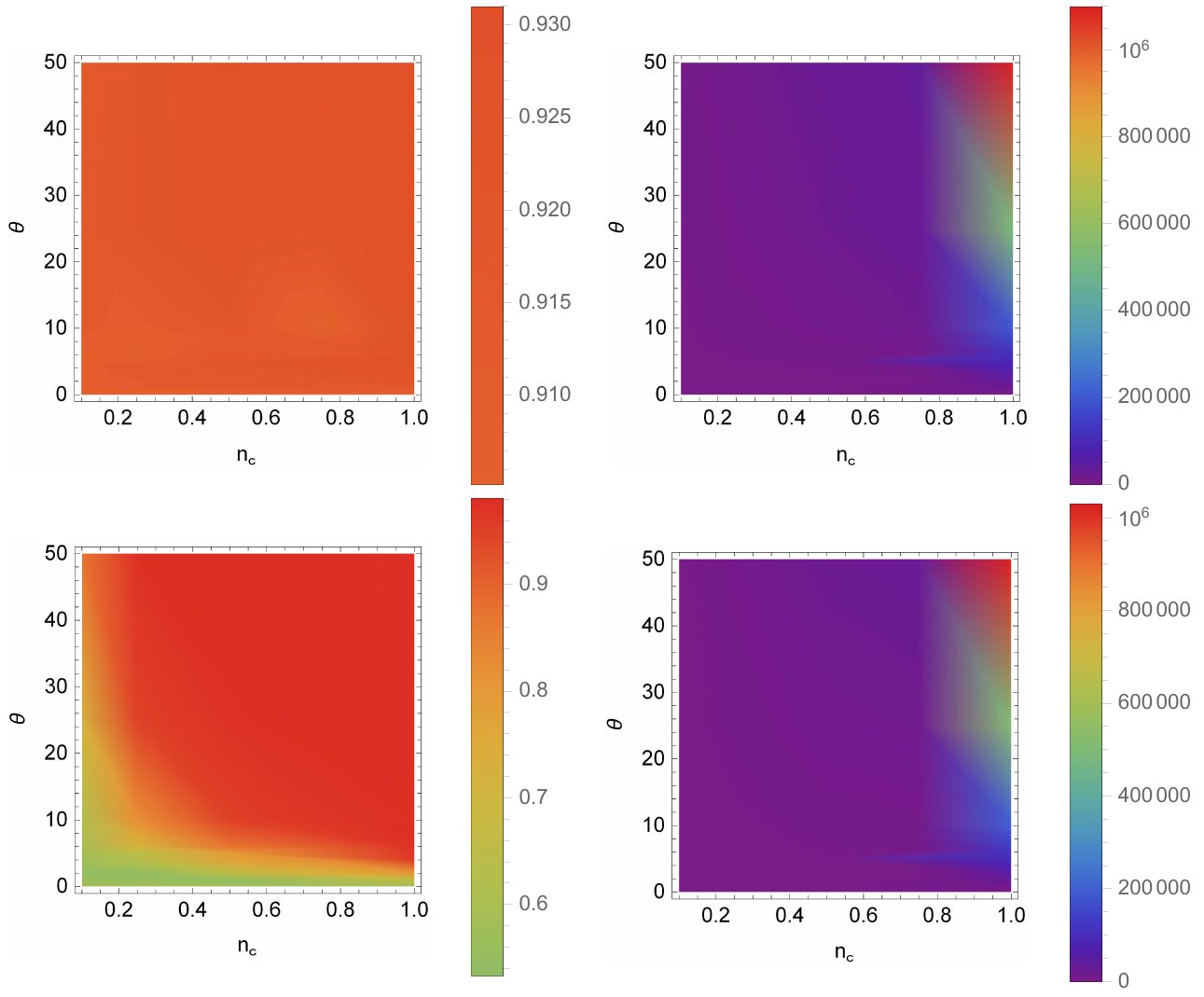}
\caption{\textbf{Combination between node influence and local cooperation based interference for BA model (top row) vs DMS model (bottom row),} for varying per-individual cost of investment $\theta$ and threshold of neighbourhood cooperation $n_C$, for $c_I = 0.05$. The left column reports the frequency of cooperation while the right one reports the total cost required. Parameters: $b = 1.8; \ n = 5000; \ z = 4$ (average node connectivity).}
\label{fig:lc-combination}
\vspace{-0.5cm}
\end{figure}

Due to the ease of acquiring information related to node connectivity in some types of networks, we test a combination of the two mechanisms where a cooperator node receives the individual investment only if both thresholds, local cooperation $n_C$ and node influence $c_I$ are met. Our results show that this is a risk-averse interference mechanism for low values of $c_I$ (See Figure ~\ref{fig:lc-combination}).

For the BA model, the possibility of inappropriate interference which leads to cyclic patterns is virtually eliminated even for very low values of $c_I$. We note that for very high values of per-individual investment $\theta$, there is a marked increase in levels of cooperation while maintaining cost efficiency, if a high enough value for $n_C$ is selected, similarly to our results for the LC-based mechanism. In that case, introducing the added parameter of node influence serves no purpose and reduces perceived gains to cooperation. 

In the case of the SF NOC with a high prevalence of triangular motifs (DMS), the combination mechanism produces similar results to the ones observed for solely LC-based interference, but with slightly reduced costs across the range of parameter values and with a more predictable correlation between threshold values, per-individual investment cost and gains in cooperation. The maximum gains to levels of cooperation are reduced slightly when compared to the single two mechanisms, with the exception of very high $\theta$. 

Following these findings, we have shown that an integrated approach to interference would work best when the nature of the network is ambiguous. In that case, this type of interference would promote converge to cooperation in the case of the DMS model, without risking the decrease in cooperation seen in the BA Model for the two interference mechanisms applied independently. 



\section{Conclusions and Future Work}

In summary, this paper aims to determine how best an external decision maker could incentivise a population of autonomous agents facing a cooperative dilemma to fulfil a coveted collective state. We build on a previous account which identified the most effective mechanisms to foster cooperative scenarios in spatially distributed systems in regular graph structured populations of agents, but instead we consider two popular models of scale-free networks of contacts. In particular, we try to understand if the insights set out in the context of regular graphs remain applicable to heterogeneous models, as well as exploring an additional avenue of interference enabled by the variance in node connectivity. To address these issues, we have combined an evolutionary game theoretic model with several incentive mechanisms in two types of pre-generated networks characterised by preferential attachment, with different clustering coefficients. We argue that this problem cannot be solved trivially and we show that transitivity (i.e. the global clustering coefficient) should be the driving force behind the choice of an interference mechanism in promoting cooperation in heterogeneous network structures, as well as its application. 

Our comparison between the two types of SF networks provides valuable insights regarding the importance of clustering in the outcome of cooperation. We found that a large clustering coefficient allows for successful, cost-effective interference, indeed even when disregarding a full comprehension of the population and its tendencies. These results are of particular interest, given that most SF networks portray high clustering, such as in the case of social ties where friends are likely to be friends of each other \citep{newman2018networks}. Moreover, heterogeneous scenarios inhibited by spatial constraints (e.g. in highly urbanised areas or even the allotment of rangelands such as pastures) also impose some measure of clustering. 

In the absence of clustering, we found that impetuously rewarding cooperators can lead to cyclic patterns which damage cooperation in the long run, and we show how this can be avoided when a decision maker lacks information about the level of clustering of the network. We observe a large negative impact on the cost of rewarding cooperators in the case of a low clustering coefficient, and provide insights on how it can be reduced. Moreover, we show that ignoring lowly connected individuals leads to unprofitable and even futile intervention irrespective of network transitivity. 

Our future work  aims to provide a comprehensive exploration of external interference  on multiple types of networks while adopting different strategy update forms, such as stochastic learning \citep{szabo2007evolutionary}. We envisage that stochasticity will increase the overall cooperation and reduce the occurrence of cyclic patterns due to reckless interference, or eliminate them altogether. Furthermore, we plan to examine spatially-motivated interference mechanisms for heterogeneous networks, encouraging the formation of links between nodes or on the contrary, cutting off said links. The inherently high levels of cooperation in heterogeneous networks motivate us to experiment with a higher bias towards defection or mechanisms specifically aimed at lowly connected nodes. 

\section{Acknowledgements}

This work was supported by   Future of Life Institute (grant RFP2-154) and {by FCT-Portugal (grants UID/CEC/50021/2013, PTDC/EEI-SII/5081/2014, and PTDC/MAT/STA/3358/2014)}.

\footnotesize
\bibliographystyle{apalike}
\bibliography{bibliography} 

\end{document}